\documentclass[usenatbib,usedcolumn]{mn2e}
\usepackage{psfig}

\newcolumntype{d}[1]{D{.}{.}{#1}}
\def\lapp{\ifmmode\stackrel{<}{_{\sim}}\else$\stackrel{<}{_{\sim}}$\fi}

\title[Pulsar Nulling and Mode Changing]{Pulsar Nulling and Mode
  Changing}

\author[N. Wang, R. N. Manchester and S. Johnston]{N. Wang$^{1,2,3}$
\thanks{E-mail:
na.wang@ms.xjb.ac.cn}, R. N. Manchester$^2$ and S. Johnston$^{2,3}$\\
$^1$Urumqi Observatory, NAOC, China\\
$^2$Australia Telescope National Facility, CSIRO, Australia\\
$^3$School of Physics, University of Sydney, Australia}

\begin{document}

\date{29 April 2006}

\maketitle

\begin{abstract}
With its relatively long observation time per pointing, the Parkes
multibeam survey was effective in detecting nulling pulsars. We have
made 2-hour observations of 23 pulsars which showed evidence for pulse
nulling or mode changing in the survey data. Because of the low flux
density of these pulsars, in most cases averaging times of between
10~s and 60~s were necessary and so this analysis is insensitive to
very short nulls. Seven of the pulsars had null fractions of more than
40\% with the largest having a lower limit of 95\%. Mode changes were
observed in six pulsars with clear relationships between nulling and
mode changing in some cases. Combined with earlier results, the data
suggest that large null fractions are more related to large
characteristic age than to long pulse period. The observations suggest
that nulling and mode changing are different manifestations of the
same phenomenon.
\end{abstract}


\section{Introduction}
Observed pulsar radio flux densities vary on many timescales. Some of
these variations result from scintillation due to scattering of ray
paths by fluctuations in the interstellar electron density
\citep{ric90}. Diffractive scintillation occurs on relatively short
timescales, typically minutes to hours, but can usually be avoided by
averaging over wide bandwidths. At least when averaged over timescales
of many minutes, the intrinsic luminosity of most pulsars is
apparently very stable \citep{ks92}. However, on shorter timescales, a
wide variety of fluctuation phenomena are observed. In all pulsars,
the pulse energy fluctuates from pulse to pulse, with modulation
indices often greater than unity \citep{tmh75,wes06}. In some pulsars,
these modulations are quasi-periodic, often only affecting part of the
pulse profile \citep[e.g.,][]{bac73}. These fluctuations appear to be
closely related to the drifting subpulses seen in some pulsars
\citep[e.g.,][]{tmh75}.

All of these modulations are basically continuous, that is, in a given
pulsar they are present most or all of the time. Pulsars also exhibit
a different class of intensity modulation which is characterised by
discontinuous transitions between otherwise stable modes. Pulse
nulling is a phenomenon in which the pulse energy suddenly drops to
zero or near zero and then just as suddenly returns to its normal
state \citep{bac70}. Nulling is relatively common in pulsars. Based on
data for 72 well-observed pulsars, \citet{big92a} found evidence for
nulling in 43. Nulls vary widely in duration, from just one or two
pulses to many hours or even days, and in ``null fraction'' (NF), the
fraction of time that the pulsar is in a null state, which can range
from zero (for the Vela pulsar) to more than 50\%. \citet{rit76} found
that the NF was correlated with pulsar age, with older pulsars more
likely to null, but \citet{big92a} found a better correlation with
pulse period. Recently \citet{klo+06} showed that in PSR B1931+24,
which has long-duration quasi-periodic nulls with a cycle time of
about 40 days, the slow-down rate is reduced by about
one third when the pulsar is in a null state, demonstrating that the
magnetospheric currents responsible for the pulse emission also
contribute to the pulsar braking. 

Mode changing is another kind of discontinuous change where the mean
pulse profile abruptly changes between two (or sometimes more)
quasi-stable states. It was first observed in PSR B1237+25, a
five-component pulsar in which pulse power occasionally switches from
the trailing two components to the central component
\citep{bac70a}. Subsequent observations have revealed mode changing in
a dozen or so pulsars, most of which have multi-component
profiles. Many of these pulsars also exhibit drifting subpulses and
nulling \citep{vkr+02,jl04,rwr05}. Drifting, quasi-periodic
modulation, microstructure and polarisation properties are all
affected by mode changing \citep{tmh75,bmsh82,ran86,gjk+94} showing
that it represents a fundamental change in the emission process. Both
nulling and mode changing are broad-band and they may be different
manifestations of the same basic phenomenon. Observations of pulse
nulling are often limited by signal-to-noise ratio and in practice a
null is only an absence of detectable emission, with the best limits
$\lapp 1$\% of the mean intensity \citep{big92a,klo+06}. For PSR
B0826$-$34, averaging of data within apparent nulls has revealed weak
emission within the ``null'' intervals at a level of $\sim 2$\% of the
mean ``on'' flux density \citep{elg+05}. Interestingly, the pulse
profile is different to that observed when the pulse is strong,
showing that the effect is really a mode change.

In this paper we present observations of 23 mostly southern pulsars,
made using the Parkes 64-m radio telescope at 1500 MHz. These pulsars,
many of which were discovered in the Parkes multibeam pulsar survey
\citep{mlc+01}, have no previously published pulse fluctuation
data. In \S 2 we describe the observations and our results and in \S 3
we discuss the implications of the results and give our conclusions.

\section{Observations and Results}
Observations were made in two sessions, 2004 March 20 -- 21 and 2004
June 8 -- 9, using the H-OH receiver on the Parkes 64-m radio
telescope. A filterbank system having a total bandwidth of 576 MHz
centred at 1518 MHz with 192 3-MHz channels on each of two orthogonal
linear polarisations was used. Signals from the two polarisations were
detected, summed, high-pass filtered, one-bit digitised at 250 $\mu$s
intervals and then written to digital linear tapes. The system
equivalent flux density on cold sky was approximately 45
Jy. 

Selection of the pulsar sample was based on data obtained in the
Parkes Multibeam Pulsar Survey. Phase-time diagrams
\citep[see][]{mlc+01} for all pulsars detected in this survey were
examined and those showing evidence for nulling were selected. Since
the time resolution of the phase-time diagrams was typically 16 seconds,
pulsars with nulls only of shorter or comparable duration were not
selected. Sensitivity limitations also meant that nulls could only be
detected among the stronger multibeam pulsars. Table~\ref{tb:obs}
lists the selected pulsars along with their basic properties in the
first seven columns: J2000 name, B1950 name (if it exists), pulse
period $P$, period first time-derivative $\dot P$, characteristic age
$\tau_c = P/(2 \dot P)$, dispersion measure (DM) and mean flux density
at 1400 MHz. For most of the pulsars listed in Table~\ref{tb:obs}, a
continuous 2-hour observation was obtained.

In off-line processing the data were dedispersed and then folded at
the topocentric pulsar period to form a series of
sub-integrations. Sub-integration lengths are given in column (8) of
Table~\ref{tb:obs}; these ranged from 10~s to 60~s depending on the
pulsar strength. This analysis is of course insensitive to nulls of
duration less than or comparable to the sub-integration length. For
two cases, PSR J1326$-$6700 and PSR J1401$-$6357, the signal/noise
ratio was sufficiently good that sub-integration averaging was not
required and intensities were computed for individual pulses. All of
these pulsars have large DMs; consequently characteristic bandwidths
for interstellar diffractive scintillation are much less than our
observed bandwidth and scintillation fluctuations do not significantly
affect the computed pulse intensities.  For example, at a DM of 100
cm$^{-3}$ pc (roughly the smallest in the sample), the typical
bandwidth for diffractive scintillation at 1400 MHz is around 50 kHz,
much less than our observed bandwidth.

\begin{figure}
\centerline{\psfig{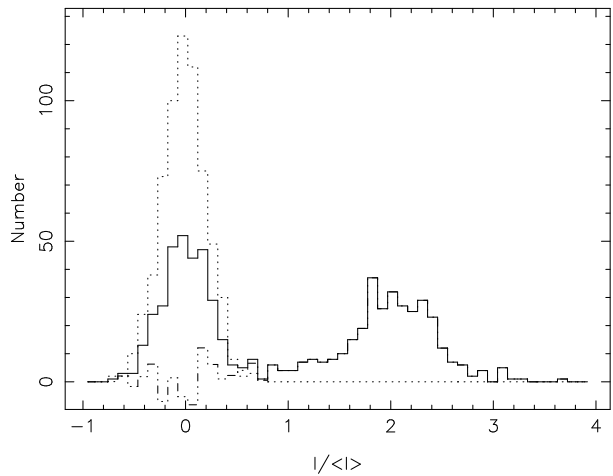}}
\caption{Histogram of on-pulse (solid line) and off-pulse (dotted
  line) intensities normalised by the mean pulse intensity for PSR
  J1049$-$5833. The dot-dashed line is the on-pulse histogram after
  subtraction of the null pulses (see text).}\label{fg:histo}
\end{figure}

Pulse energies for each sub-integration were determined by integrating
under the pulse after subtraction of an off-pulse baseline and
off-pulse energies were computed from a window of the same duration in
the baseline region. Null fractions (NF), listed in column (9) of
Table~\ref{tb:obs}, were determined as follows. Histograms of the
on-pulse and off-pulse intensities were formed (Fig.~\ref{fg:histo})
both of which sum to the total number of sub-integrations $N$. A
scaled version of the off-pulse histogram was subtracted from the
on-pulse histogram so that the sum of the difference counts in
bins with $I<0$ was zero. The NF is then simply the scale factor with
an uncertainty $\sqrt{n_p}/N$, where $n_p$ is the number of null
sub-integrations. Null lengths were measured as the time between the
first pulse energy below a threshold (typically five times the rms
fluctuation in the off-pulse energies) to the next above the
threshold; column (10) lists the mean and rms values of these lengths
for each observation. The final column lists the mean and rms null
cycle times, that is, the mean and rms times from the onset of one
null to the onset of the next.

In most cases, the rms scatter of the null length and null cycle time
are comparable to the mean value, indicating a relatively ``white''
fluctuation spectrum. This was verified by directly computing power
spectra for the sequence of pulse energies using a Fast Fourier
Transform routine. However, in a few cases, e.g., PSR J1920+1040,
quasi-periodic fluctuations are seen and the rms scatter of the cycle
time is significantly less than the mean value. Because of the need to
integrate for 10~s or more to give sufficient signal to noise for
these relatively weak pulsars, this analysis was insensitive to
periodic fluctuations related to subpulse drifting \citep[e.g.,][]{wes06}.

\begin{table*}
\centering
\begin{minipage}{140mm}
\caption{Observed pulsars and nulling parameters}\label{tb:obs}
\footnotesize
\begin{tabular}{lld{3}d{3}d{3}d{2}d{2}d{2}rll}
\hline 
\multicolumn{1}{c}{(1)} & \multicolumn{1}{c}{(2)} & \multicolumn{1}{c}{(3)} & 
  \multicolumn{1}{c}{(4)} & \multicolumn{1}{c}{(5)} & \multicolumn{1}{c}{(6)} & 
  \multicolumn{1}{c}{(7)} & \multicolumn{1}{c}{(8)} & \multicolumn{1}{c}{(9)} & 
  \multicolumn{1}{c}{(10)} & \multicolumn{1}{c}{(11)} \\
\multicolumn{1}{c}{J2000} &\multicolumn{1}{c}{B1950} &\multicolumn{1}{c}{$P$} &
  \multicolumn{1}{c}{$\dot P$} & \multicolumn{1}{c}{Age} & \multicolumn{1}{c}{DM} &  
  \multicolumn{1}{c}{$S_{1400}$} & \multicolumn{1}{c}{Int.} & NF &
  Null len. &  Null cycle  \\
\multicolumn{1}{c}{Name} & \multicolumn{1}{c}{Name} & \multicolumn{1}{c}{(s)} &
  \multicolumn{1}{c}{($10^{-15}$)} & \multicolumn{1}{c}{($10^6$ yr)} & 
  \multicolumn{1}{c}{(cm$^{-3}$ pc)} & \multicolumn{1}{c}{(mJy)} & 
  \multicolumn{1}{c}{(s)} & (\%) & \multicolumn{1}{c}{(s)} & \multicolumn{1}{c}{(s)} \\
\hline
J0846$-$3533  & B0844$-$35 & 1.116 &  1.60 & 11.0 &   94 & 2.7  & 10 &   0    &  ...       &    ...       \\
J1032$-$5911  & B1030$-$58 & 0.464 &  3.00 &  2.4 &  418 & 0.93 & 20 &   0    &  ...       &    ...       \\
J1049$-$5833  & ...      & 2.202 &  4.41 &  7.9 &  447 & 0.72 & 10 & 47(3)  &  84(86)    &   179(127)   \\
J1326$-$6700  & B1322$-$66 & 0.543 &  5.31 &  1.6 &  210 & 11.0 & 0  & 9.1(3) &  1(1)      &   14(32)     \\
J1401$-$6357  & B1358$-$63 & 0.842 & 16.70 &  0.8 &   98 & 6.2  & 0  & 1.6(1) &  2(3)      &   120(95)   \\ \\
J1412$-$6145  & ...      & 0.315 & 98.70 &  0.1 &  515 & 0.47 & 20 &  0     &  ...       &    ...       \\
J1502$-$5653  & ...      & 0.535 &  1.83 &  4.6 &  194 & 0.39 & 10 & 93(4)  &  450(335)  &   515(360)   \\
J1525$-$5417  & ...      & 1.011 & 16.20 &  1.0 &  235 & 0.18 & 20 & 16(5)  &  28(25)    &   138(102)   \\
J1648$-$4458  & ...      & 0.629 &  1.85 &  5.4 &  925 & 0.55 & 60 & 1.4(11)&  ...       &   ...        \\
J1658$-$4306  & ...      & 1.166 & 42.80 &  0.4 &  845 & 0.80 & 60 &  0     &  ...       &   ...        \\ \\
J1701$-$3726  & B1658$-$37 & 2.454 & 11.10 &  3.5 &  303 & 2.9  & 10 & 14(2)  &  16(9)     &   118(91)    \\
J1702$-$4428  & ...      & 2.123 &  3.30 & 10.2 &  395 & 0.38 & 30 & 26(3)  &  42(24)    &   149(83)    \\
J1703$-$4851  & ...      & 1.396 &  5.08 &  4.3 &  150 & 1.10 & 10 & 1.1(4) &  ...       &    ...       \\
J1717$-$4054  & B1713$-$40 & 0.887 &   ... &  ... &  307 & 54   & 10 &  $>95$ & $>7000$    &   $>7000$    \\
J1722$-$3632  & B1718$-$36 & 0.399 &  4.46 &  1.4 &  416 & 1.60 & 20 &   0    &  ...       &    ...       \\ \\
J1727$-$2739  & ...      & 1.293 &  1.10 & 18.6 &  147 & 1.60 & 10 & 52(3)  &  48(37)    &   91(58)     \\
J1812$-$1718  & B1809$-$173& 1.205 & 19.10 &  1.0 &  254 & 1.00 & 20 & 5.8(4) &   23(6)    &   359(303)   \\
J1820$-$0509  & ...      & 0.337 &  0.93 &  5.7 &  104 & *    & 10 & 67(3)  &   71(69)   &   104(68)    \\
J1831$-$1223  & ...      & 2.857 &  5.47 &  8.3 &  342 & 1.2  & 20 & 4(1)   &   23(7)    &   448(348)   \\
J1833$-$1055  & ...      & 0.633 &  0.53 & 19.1 &  543 & 0.5  & 30 & 7(2)   &   ...      &   ...        \\ \\
J1843$-$0211  & ...      & 2.027 & 14.40 &  2.2 &  442 & 0.93 & 30 & 6(2)   &  46(23)    &   826(872)   \\
J1916+1023    & ...      & 1.618 &  0.68 & 37.7 &  330 & 0.36 & 60 & 47(4)  &  70(43)    &   152(53)    \\
J1920+1040    & ...      & 2.215 &  6.48 &  5.4 &  304 & 0.57 & 10 & 50(4)  &   43(25)   &    85(35)    \\
\hline
\end{tabular}
\end{minipage}
\end{table*}

\begin{figure*}
\centerline{\psfig{file=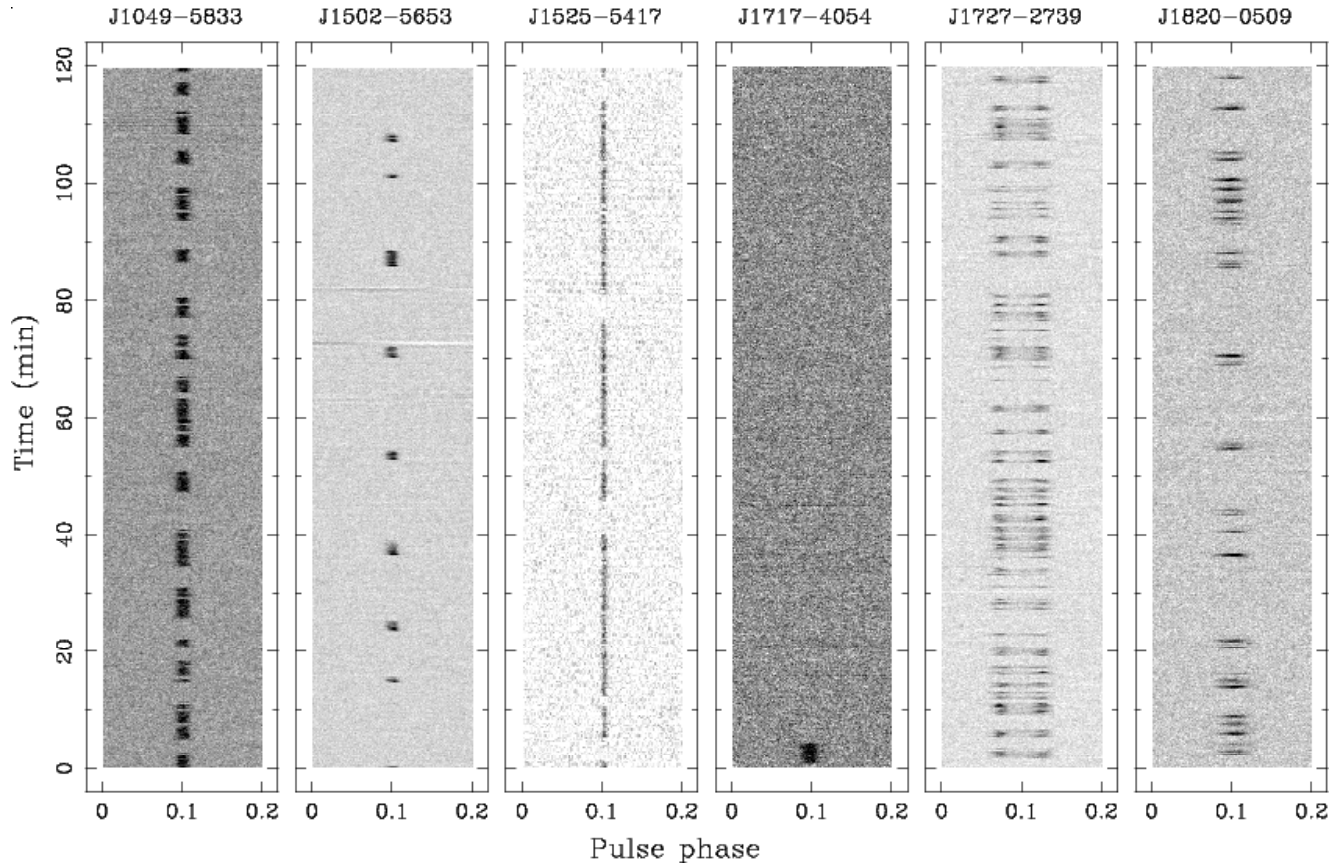,width=175mm}}
\caption{Phase-time plots for six pulsars which appear to show simple
  nulling. One fifth of the pulse period is shown in each case and the
  greyscale is linear in intensity from zero (white) to
  the maximum observed value (black).}\label{fg:nulls}
\end{figure*}

Fig.~\ref{fg:nulls} shows the observed pulse energy fluctuations as
a phase-time greyscale plot for six pulsars which appear to exhibit
simple nulls, that is, a cessation of pulse emission with no
associated mode changing. Observed NFs cover a wide range, with some
pulsars, e.g., PSR J1525$-$5417, on most of the time, to nearly 100\%
for PSR J1717$-$4054. In this latter case, only one short burst was
observed in the whole 2 hours, so the true null fraction is quite
uncertain. Observed null durations and null cycle times also vary
widely with some pulsars, e.g., PSR J1727$-$2739, having frequent
short nulls and others having long timescales.

\begin{figure*}
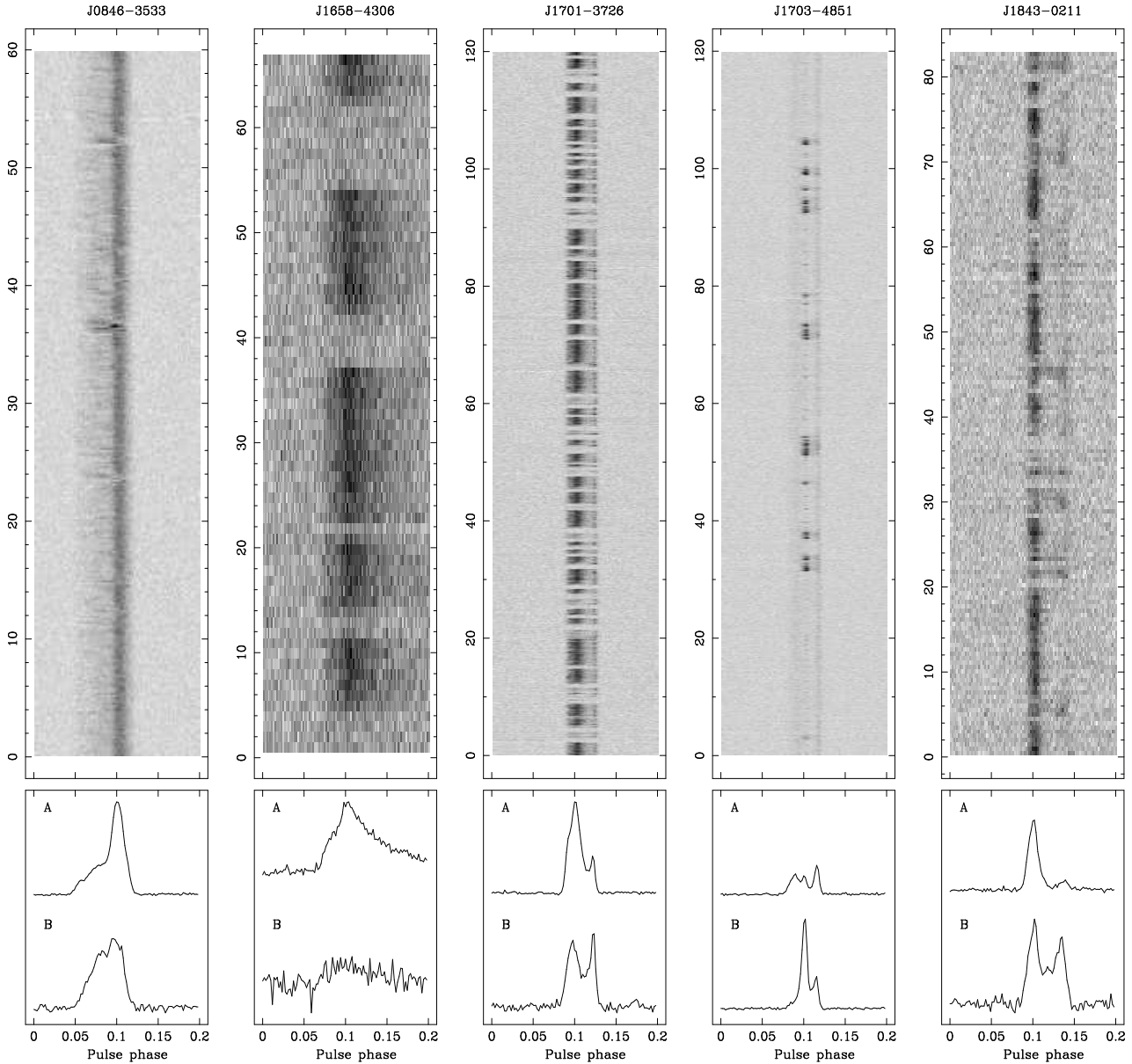

\begin{center} 
\begin{tabular}{ccccc}
\mbox{\psfig{file=0846_null_prf_nr.ps,height=160mm,angle=270}} &
\mbox{\psfig{file=1658_null_prf_nr.ps,height=160mm,angle=270}} &
\mbox{\psfig{file=1701_null_prf_nr.ps,height=160mm,angle=270}} &
\mbox{\psfig{file=1703_null_prf_nr.ps,height=160mm,angle=270}} &
\mbox{\psfig{file=1843_null_prf_nr.ps,height=160mm,angle=270}} 
\end{tabular}
\caption{Phase-time plots for five pulsars which show mode
  changing. The greyscale is linear in intensity from zero to the
  maximum observed value. In the lower part of the figure, mean pulse
  profiles for the two observed modes are shown.}
\label{fg:mode}
\end{center}
\end{figure*}

Fig.~\ref{fg:mode} collects together a group of five pulsars which
show clear mode changes. In each case there are two identifiable
modes. Mode A is defined to be the more common mode whereas mode B is
less frequently observed. In four of these pulsars, the mean flux
density is less in mode B, but for PSR J1703$-$4851 mode B is much
stronger than mode A and dominates the long-term mean profile even
though it is present for only about 15\% of the time. For
PSR J1701$-$3726 and PSR J1843$-$0211 the different modes are
separated by null periods, but this is not so for the other three
pulsars.

In the remainder of this Section we briefly discuss each pulsar in
turn.

\medskip\noindent
{\bf PSR J0846$-$3533 (B0844$-$35)} 

The phase-time plot shown in Fig.~\ref{fg:mode} shows that this
apparently 3-component pulsar has brief intervals where it changes to
a second mode (B) characterised by weaker emission in the leading and
trailing components and stronger emission in the central
component. These intervals, typically just 5 - 10 pulses long, are
usually preceded by a few very weak (but not null) pulses. In fact no
real nulls were observed in the 2-hour observation. As observed in
other pulsars \citep{bac73,ran86}, the intensity of the central
component fluctuates to a much greater extent giving it a fluctuation
spectrum which is ``redder'' than that for the other two components.

\medskip\noindent
{\bf PSR J1032$-$5911 (B1030$-$58)}

This pulsar has a simple, single-component profile of 50\% width
7.7~ms \citep{hfs+04} or just 0.017 of the period. Pulse energies are
very highly modulated with strong bursts of emission 1 -- 3 min in
duration separated by similarly short intervals of weaker
emission. There is no convincing evidence for real nulls in this
pulsar.

\medskip\noindent
{\bf PSR J1049$-$5833}

This long-period pulsar, discovered in the Parkes multibeam survey
\citep{mlc+01}, also has a simple, single-component profile. The
phase-time diagram given in Fig.~\ref{fg:nulls} shows it to have
frequent nulls with a typical duration of about 2~min. The active
intervals are of similar duration giving it a NF of close to 50\%
(Table~\ref{tb:obs}). The modulation spectrum has excess power at
periods between 250~s and 750~s, somewhat longer than the mean cycle
time of 179~s (Table~\ref{tb:obs}). This difference can be attributed
to the presence of numerous short nulls which interrupt longer bursts
of emission (Fig.~\ref{fg:nulls}). These will reduce the computed
mean cycle time, but will just add white noise to the power spectrum.

\medskip\noindent
{\bf PSR J1326$-$6700 (B1322$-$66)}

\begin{figure}
\centerline{\psfig{file=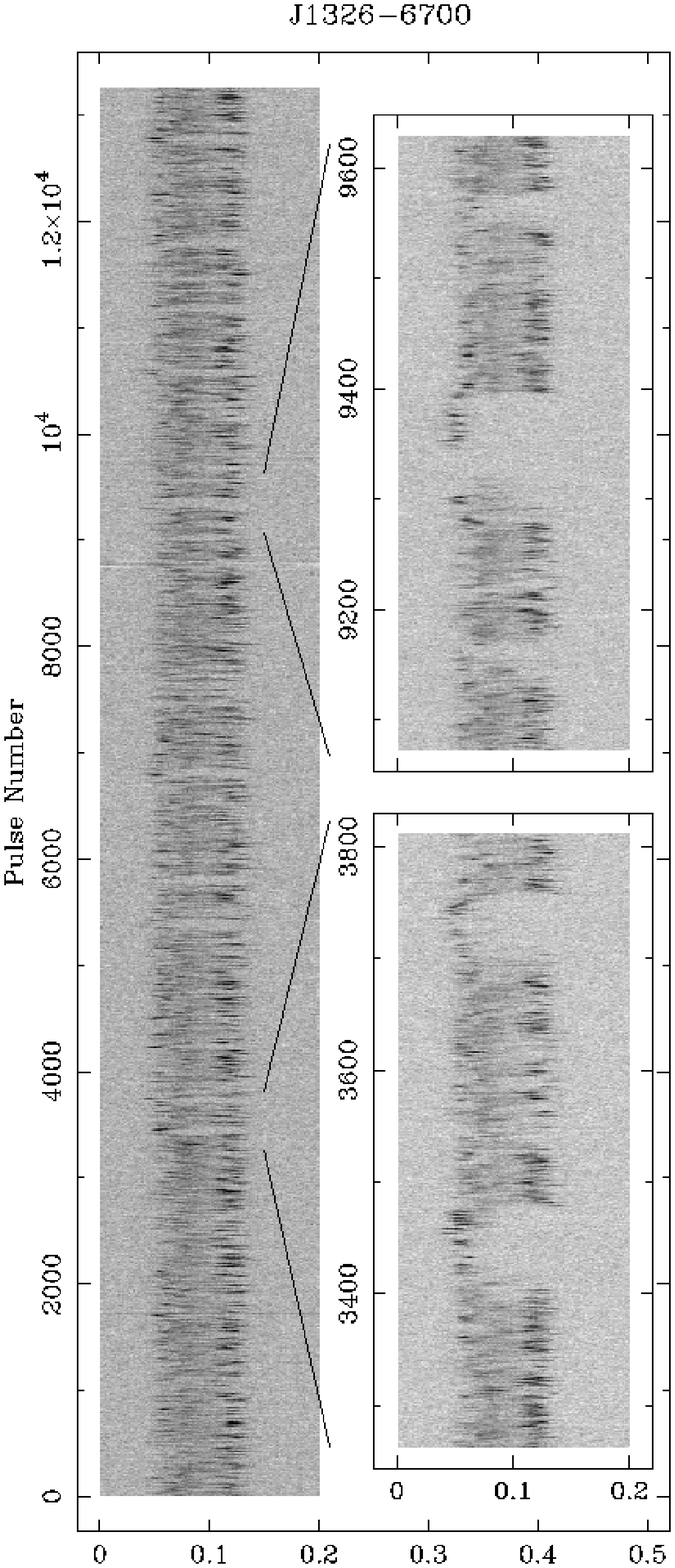,width=80mm}}
\caption{Phase-time diagram showing intensity variations for
  individual pulses from PSR J1326$-$6700. The insets show two
  portions of the data with an expanded scale.}\label{fg:1326}
\end{figure}

\begin{figure}
\centerline{\psfig{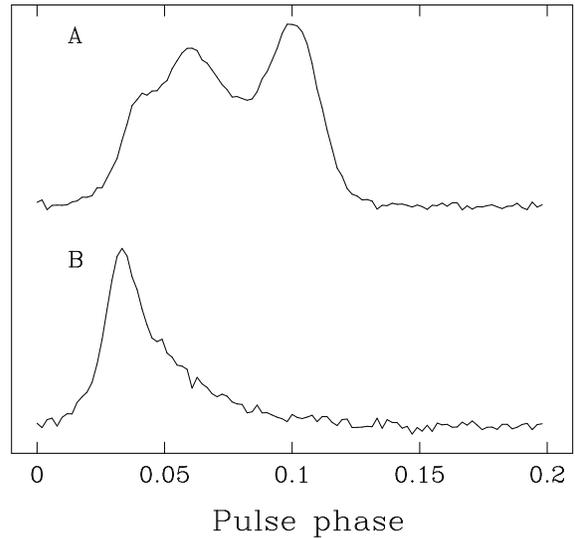}}
\caption{Mean pulse profiles for the two modes of emission from PSR
  J1326$-$6700. }\label{fg:1326mode}
\end{figure}

This pulsar has a mean pulse profile consisting of a basically triple
profile with a weaker component on the leading edge of the profile. As
shown in Fig.~\ref{fg:1326}, it displays an intriguing combination of
nulling and mode changing. Since it is relatively strong, averaging of
data into subintegrations was not required and Fig.~\ref{fg:1326}
shows individual pulse intensities. At intervals of 2 -- 10 min (200
-- 1000 pulses) emission from the two main components ceases,
typically for less than a minute.  (Table~\ref{tb:obs}). During these
``component'' nulls, sporadic emission appears at the leading edge of
the profile, forming the fourth leading component of the profile.  The
pulse phase of this mode-B emission is somewhat variable (see, for
example, around pulse 3700 -- 3750 in the lower inset of
Fig.~\ref{fg:1326}) and there are frequent short intervals during this
mode-B emission when this leading component is undetectable. The mode
A emission is more consistent with most pulses being visible although
the relative (and absolute) amplitudes of the components vary
greatly. Fig.~\ref{fg:1326mode} shows the very different mean pulse
profiles for the two modes. There is some evidence for occasional
drifting subpulses in this pulsar, for example, around pulses 3350 and
3560 in the lower inset. The drift is toward the trailing edge with a
rate of approximately 0.008 cycles/period.
 

\medskip\noindent
{\bf PSR J1401$-$6357 (B1358$-$63)}

\begin{figure}
\centerline{\psfig{file=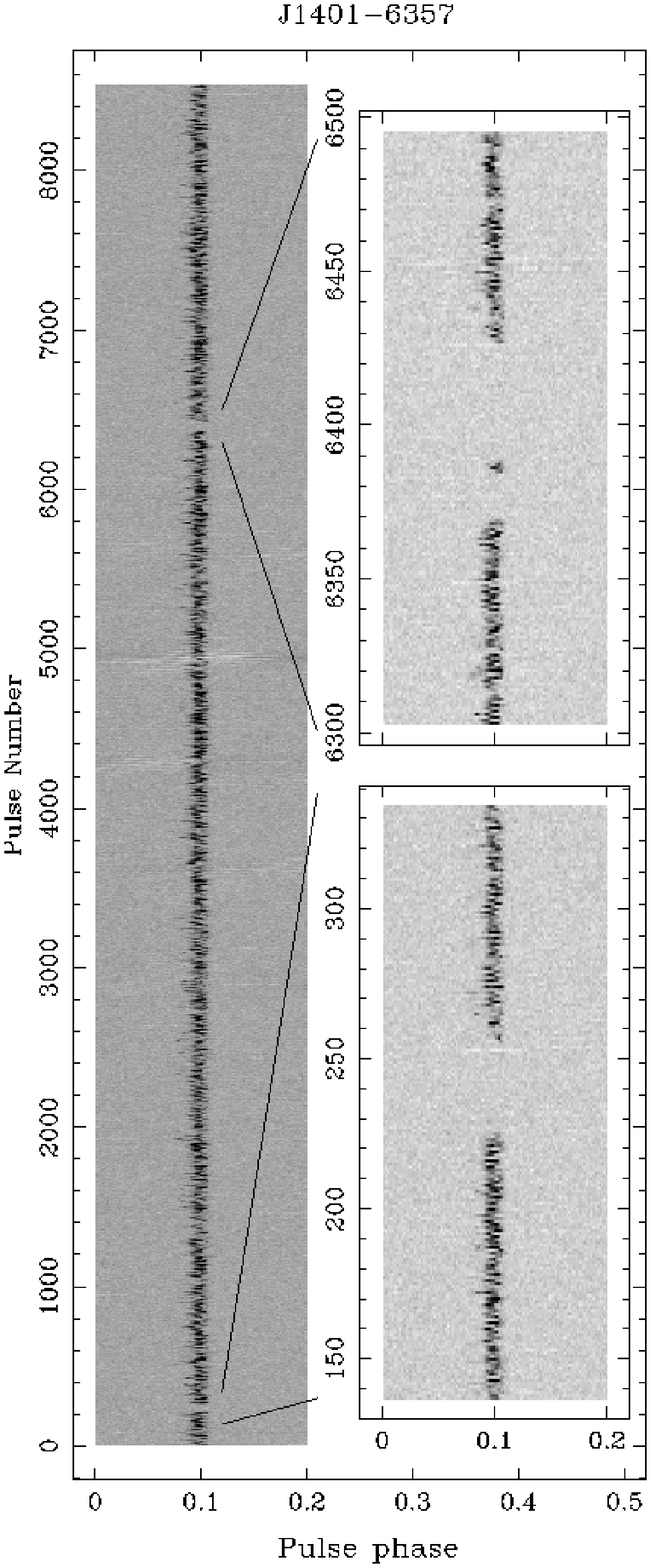,width=80mm}}
\caption{Phase-time diagram showing intensity variations for
  individual pulses from PSR J1401$-$6357. The insets show two
  portions of the data with an expanded scale. }\label{fg:1401}
\end{figure}

This pulsar has a narrow profile (50\% width of 10~ms or 0.012 in
phase) with a weak leading component \citep{hfs+04}. Individual pulse
observations shown in Fig.~\ref{fg:1401} show it to have infrequent
short nulls with an observed NF of just 1.6\%
(Table~\ref{tb:obs}). Generally the active periods are relatively long
(10 -- 30 min), but as the upper inset in Fig.~\ref{fg:1401} shows,
there can also be very short bursts of emission within a null
interval. Fig.~\ref{fg:1401} also shows that the leading component is
weak in the average profile because it is highly sporadic, with just
occasional pulses which are comparable in flux density to those at the
main peak.

\medskip\noindent {\bf PSR J1412$-$6145} This relatively weak pulsar
showed no evidence for nulls in a 2-hr observation.

\medskip\noindent
{\bf PSR J1502$-$5653}

This pulsar has a NF of 93\% (Table~\ref{tb:obs}), among the
highest known. The phase-time diagram given in Fig.~\ref{fg:nulls}
shows that active intervals last typically just a minute or so and are
separated by 10 -- 15 min. Consistent with these results, the
fluctuation power spectrum shows a broad peak at periods between 400
and 1000~s. Although difficult to see in Fig.~\ref{fg:nulls}, closer
examination of the data show that there are often short nulls of
duration 5 -- 10 periods within the bursts.

\medskip\noindent
{\bf PSR J1525$-$5417} 

This weak pulsar has a very narrow pulse of 50\% width 15~ms or 0.015
in phase \citep{kbm+03}. Observations for this pulsar were affected by
strong out-of-band interference, causing fluctuations in the baseline
level, but the phase-time diagram (Fig.~\ref{fg:nulls}) shows it to
have nulls of typical duration 3 -- 4 min. With a NF of about 16\%
(Table~\ref{tb:obs}), the active intervals are typically $\sim 20$ min
long but Fig.~\ref{fg:nulls} shows that some bursts are much
shorter. The fluctuation power spectrum has a broad peak at about
2400~s.

\medskip\noindent
{\bf PSR J1648$-$4458}

Another relatively weak pulsar from the Parkes multibeam survey
\citep{kbm+03}, this pulsar has sporadic bursts of emission with burst
durations ranging between a few minutes and 15~min and a mean duty
cycle of about 50\%. However, weaker emission is clearly visible
between the bursts, so there are no true nulls. Although there are no
obvious differences in the mean pulse profiles for the burst and weak
intervals, this may a form of mode changing.

\medskip\noindent
{\bf PSR J1658$-$4306}

This pulsar has a high DM and the pulse profile is significantly
affected by interstellar scattering \citep{kbm+03}. It is quite weak
and 60~s integrations were necessary to give adequate signal-to-noise
ratio. Fig.~\ref{fg:mode} appears to show five null periods of average
duration 3 -- 4 min in the 70-min data span. However, closer
examination of the figure shows that, like PSR J1648$-$4458, there is
weak but significant emission in the ``null'' intervals. This is
verified by the pulse energy histogram which is bimodal but shows no
null pulses. The apparent nulling is therefore better described as
mode changing. Because of the low signal/noise ratio of the mode B
profile, it is difficult to tell if the profiles for the two modes are
different. The intrinsic (unscattered) mode A profile is probably
double and there is some indication that the stronger trailing
component is absent or weaker in the mode B profile.

\medskip\noindent
{\bf PSR J1701$-$3726 (B1658$-$37)}

As Fig.~\ref{fg:mode} shows, this pulsar has frequent short nulls
giving a broad peak in the fluctuation spectrum for periods between 60
and 500~s. Most of the time, the pulsar has a three-component profile
with the middle (but not central) component the strongest. However, as
illustrated in Fig.~\ref{fg:mode}, on occasions this middle component
becomes much weaker and the pulse profile is basically double. Within
a given burst, the pulsar remains in one mode, i.e., the two modes are
always separated by a null interval. Mode B emission is both rarer and
weaker than mode A, with only seven or eight short bursts, each of a
minute or two in duration, during the 2-hr observation.

\medskip\noindent
{\bf PSR J1702$-$4428}

Another long-period and rather weak pulsar which fluctuates greatly in
strength; it appears to be in a null state about one quarter of the
time. The fluctuation spectrum has excess power in the range 85~s to
1500~s.

\medskip\noindent
{\bf PSR J1703$-$4851}

As shown in Fig.~\ref{fg:mode}, this pulsar exhibits a fascinating
form of mode changing in which the central and trailing components of
the triple profile are intermittently enhanced relative to the leading
component. We define the most common, although in this case weaker,
mode to be mode A. Mode A emission appears to be quite steady, at
least with the 10-s integrations used in Fig.~\ref{fg:mode}. The
enhanced emission of mode B occurs in short bursts of duration from
just a few seconds up to 1 or 2 min and is present for about 15\% of
the time. For at least the central component, the longer bursts of
mode B emission are interrupted by short intervals of weaker or maybe
no emission.

\medskip\noindent
{\bf PSR J1717$-$4054 (B1713$-$40)} 

This pulsar had only one intense burst of duration about 3.5 min
during the whole 2-hr observation (Fig.~\ref{fg:nulls}), giving it a
null fraction of at least 95\%, the highest value measured so
far. Although discovered in the Johnston et al. (1992)\nocite{jlm+92}
Galactic Plane survey, this pulsar still has no measured $\dot P$ and
hence characteristic age. This is certainly because the high
NF makes it difficult to observe. We note that the lack of a measured
$\dot P$ for this strongly nulling pulsar is likely to bias statistics
on the dependence of nulling on $\dot P$-related parameters.

\medskip\noindent
{\bf PSR J1722$-$3631 (B1718$-$36)}

Pulse emission from this pulsar is highly modulated but no nulls were
observed in a 2-hr observation.

\medskip\noindent
{\bf PSR J1727$-$2739}

This pulsar has a classic double mean pulse profile of 50\% width
90~ms or 0.07 of the pulse period. The phase-time diagram given in
Fig.~\ref{fg:nulls} shows that the pulsar emits frequent short
bursts separated by null intervals. There is no evidence for any mode
changing. Typical burst lengths are about a minute and null lengths a
little longer, although both are quite variable, giving a rather flat
modulation spectrum. The observed NF is about 50\%
(Table~\ref{tb:obs}).

\medskip\noindent
{\bf PSR J1812$-$1718 (B1809$-$173)}

The mean pulse profile for this pulsar is dominated by a narrow
leading component of 50\% width 19~ms or 0.015 periods
\citep{hfs+04}. There is a weak trailing component of similar width
which overlaps with the main component. The phase-time diagram shows
frequent short nulls of average duration about 25~s with an observed
NF of about 6\%. There are no significant periodicities in the modulation
spectrum (with sampling interval 20~s).

\medskip\noindent 
{\bf PSR J1820$-$0509} 

This pulsar, discovered in the second ``reprocessing'' phase of the
Parkes multibeam survey \citep{lfl+06}, has a relatively short period
but is old with a characteristic age of 5.7 Myr
(Table~\ref{tb:obs}). As Fig.~\ref{fg:nulls} shows, it is in a null
state most of the time with only short active periods. The derived NF
(Table~\ref{tb:obs}) is 67\% and the mean on-time is less than a
minute. The fluctuation power spectrum shows no significant
periodicities and there is no evidence for mode changing.

\medskip\noindent 
{\bf PSR J1831$-$1223}

This long-period pulsar has a double-peaked mean profile of 50\% width
98~ms or 0.034 in pulse phase \citep{mhl+02}. It suffers frequent
short nulls of mean duration about 25~s (Table~\ref{tb:obs}) but the
NF is low. The phase-time plot suggests possible mode changing, but
unfortunately the signal-to-noise ratio is rather poor and a more
definitive statement cannot be made based on these data. The
fluctuation power spectrum has excess power at periods between 50 and
200~s.

\medskip\noindent 
{\bf PSR J1833$-$1055}

This weak pulsar required an integration time of 30~s to give adequate
signal-to-noise ratio in the phase-time plot. The pulsar is very
bursty with intervals of strong emission typically lasting a minute or
two. Weaker emission is seen between these bursts but there appear to
be a few per cent of null pulses. There is a peak in the
fluctuation power spectrum at about 400~s period with excess power
between 200 and 900~s.

\medskip\noindent 
{\bf PSR J1843$-$0211}

This long-period pulsar has a moderately large period derivative,
giving it a characteristic age of 2.2 Myr (Table~\ref{tb:obs}). The
mean pulse profile has a strong leading component and much weaker
trailing component, but is never-the-less clearly
double-peaked. Although weak (30~s integrations were required) the
pulsar appears to have occasional short nulls, of typical duration
less than or about one minute (Fig.~\ref{fg:mode}). There is clear
evidence of mode changing with the leading component much stronger in
the dominant mode (A) and the two components weaker and of
approximately equal amplitude in the secondary mode (B). Although mode
B is present for only a few per cent of the time, it is the dominant
contributor to the trailing component of the mean profile.

\medskip\noindent 
{\bf PSR J1916+1023}

This pulsar has a two- or possibly three-component profile
\citep{hfs+04}. It has a large characteristic age and is very weak,
requiring 60-s integrations for sufficient signal-to-noise ratio. It
appears to be in a null state for about 50\% of the time with many
short bursts of emission, typically just one or two minutes in
duration. There is no significant evidence for mode changing.

\medskip\noindent 
{\bf PSR J1920+1040}

This pulsar has single-component mean pulse profile of 50\% width
45~ms or 0.020 of the period \citep{hfs+04}. It has frequent
quasi-periodic bursts of typical duration about 40~s separated by
somewhat longer nulls, giving it a null fraction of about 50\%
(Table~\ref{tb:obs}). There is no evidence for mode changing. The
fluctuation power spectrum has a strong peak at about 90~s periodicity
with excess power between 60 and 340~s.

\section{Discussion}
Most of the pulsars in this study were discovered in the Parkes
multibeam pulsar survey. The observation time per pointing in this
survey, 35 min, was much longer than for other large-scale
surveys. This long integration time facilitated the discovery of
longer-period pulsars \citep[e.g.][]{msk+03} which are more likely to
null \citep{big92a}. Furthermore, it aided the
identification of nulling pulsars whose off times are more than a few
minutes, sources that would often be missed by surveys with
observation times of this order. The extreme example of this is the
discovery of ``rotating radio transients'' or RRATs \citep{mll+06}, pulsars
in which just single pulses are observed at intervals of minutes to
hours. 

\begin{figure}
\centerline{\psfig{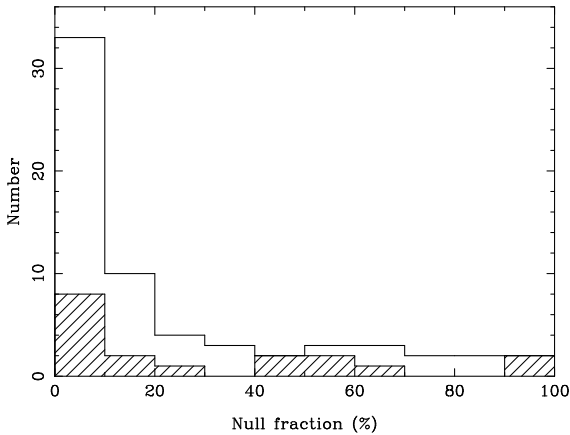}}
\caption{Histogram of observed null fractions. Values from the present
  observations are shown hatched.}\label{fg:null_hist}
\end{figure}

As can be seen from Table~\ref{tb:obs}, seven of the 23 pulsars in our
sample have NFs of 50\% or more (within the uncertainties), with two
having NFs of order 95\%, the largest known (apart from the
RRATs). Table~\ref{tb:null} lists previously known nulling pulsars and
their NF. While some of these values may be misleading because of
low signal-to-noise ratio in the observations, the table gives the
best available indication of the distribution of nulling in known
pulsars. Fig.~\ref{fg:null_hist} shows that the present work has
substantially increased the number of pulsars with high NF, doubling
the number known with NF $\ge 50$\%. Although our observations are
generally not very sensitive to short nulls (because of limited
signal/noise on these relatively weak pulsars), it is clear that
pulsars with low NF are much more common than those with high NF.

\begin{table*}
\centering
\begin{minipage}{110mm}
\caption{Previously known nulling pulsars}\label{tb:null}
\footnotesize
\begin{tabular}{lld{3}d{2}d{1}d{2}l}
\hline 
\multicolumn{1}{c}{J2000} &\multicolumn{1}{c}{B1950} &\multicolumn{1}{c}{$P$} &
  \multicolumn{1}{c}{$\dot P$} & \multicolumn{1}{c}{Age} & NF & Reference\\
\multicolumn{1}{c}{name} & \multicolumn{1}{c}{name} & \multicolumn{1}{c}{(s)} &
  \multicolumn{1}{c}{($10^{-15}$)} & \multicolumn{1}{c}{($10^6$ yr)} & (\%) & \\
\hline
J0034$-$0721 & B0031$-$07 &  0.943 &   0.41 &  36.6 &   37.7  & Huguenin, Taylor \& Troland (1970) \\
J0151$-$0635 & B0148$-$06 &  1.465 &   0.44 &  52.4 &   2.5  &  Biggs (1992)        \\
J0304+1932 & B0301+19 &  1.388 &   1.30 &  17.0 &   10    &  Rankin (1986)       \\
J0452$-$1759 & B0450$-$18 &  0.549 &   5.75 &   1.5 &   0.45 &  Biggs (1992)       \\
J0528+2200 & B0525+21 &  3.746 &  40.00 &   1.5 &   25    &  Ritchings (1976)        \\
J0630$-$2834 & B0628$-$28 &  1.244 &   7.12 &   2.8 &   0.15 &  Biggs (1992)       \\
J0659+1414 & B0656+14 &  0.385 &  55.00 &   0.1 &   12    &  Biggs (1992)        \\
J0738$-$4042 & B0736$-$40 &  0.375 &   1.62 &   3.7 &   0.2  &  Biggs (1992)        \\
J0742$-$2822 & B0740$-$28 &  0.167 &  16.80 &   0.2 &   0.1  &  Biggs (1992)        \\
J0754+3231 & B0751+32 &  1.442 &   1.08 &  21.2 &   34.0  &  Weisberg et al. (1986)       \\
J0814+7429 & B0809+74 &  1.292 &   0.17 & 122.0 &   1.42 &  Lyne \& Ashworth (1983)      \\
J0820$-$1350 & B0818$-$13 &  1.238 &   2.11 &   9.3 &  1.01 &  Lyne \& Ashworth (1983)      \\
J0828$-$3417 & B0826$-$34 &  1.849 &   1.00 &  29.4 &  80    &  Durdin et al. (1979)      \\
J0837+0610 & B0834+06 &  1.274 &   6.80 &   3.0 &   7.1  &  Ritchings (1976)        \\
J0837$-$4135 & B0835$-$41 &  0.752 &   3.54 &   3.4 &   0.6  &  Biggs (1992)        \\
J0943+1631 & B0940+16 &  1.087 &   0.09 & 189.0 &   8    &  Weisberg et al. (1986)        \\
J0942$-$5552 & B0940$-$55 &  0.664 &  22.90 &   0.5 &   6    & Biggs (1992)       \\
J1115+5030 & B1112+50 &  1.656 &   2.49 &  10.5 &   60    &  Ritchings (1976)        \\
J1136+1551 & B1133+16 &  1.188 &   3.73 &   5.0 &   15.0  &  Ritchings (1976)        \\
J1239+2453 & B1237+25 &  1.382 &   0.96 &  22.8 &   6.0  &  Ritchings (1976)        \\
J1243$-$6423 & B1240$-$64 &  0.388 &   4.50 &   1.4 &   2    &  Biggs (1992)        \\
J1430$-$6623 & B1426$-$66 &  0.785 &   2.77 &   4.5 &   0.025&  Biggs (1992)      \\
J1456$-$6843 & B1451$-$68 &  0.263 &   0.10 &  42.5 &   1.7  &  Biggs (1992)        \\
J1532+2745 & B1530+27 &  1.125 &   0.78 &  22.9 &   6    &  Weisberg et al. (1986)       \\
J1649+2533 & ...      &  1.015 &   0.56 &  28.8 &   30    & Lewandowski et al. (2004)  \\
J1731$-$4744 & B1727$-$47 &  0.830 & 164.00 &   0.1 &   0.05 &  Biggs (1992)       \\
J1744$-$3922 & ...      &  0.172 &   0.00 &1760.0 &   75    & Faulkner et al. (2004)    \\	
J1745$-$3040 & B1742$-$30 &  0.367 &  10.70 &   0.5 &   9  &  Biggs (1992)        \\
J1752$-$2806 & B1749$-$28 &  0.563 &   8.13 &   1.1 &   0.7   &  Biggs (1992)       \\
J1752+2359 & ...      &  0.409 &   0.64 &  10.1 &   75    &  Lewandowski et al. (2004)   \\
J1820$-$0427 & B1818$-$04 &  0.598 &   6.33 &   1.5 &   0.13 &  Biggs (1992)       \\
J1900$-$2600 & B1857$-$26 &  0.612 &   0.21 &  47.4 &   10.0  &  Ritchings (1976)        \\
J1910+0358 & B1907+03 &  2.330 &   4.47 &   8.3 &   4    &  Biggs (1992)        \\
J1932+1059 & B1929+10 &  0.227 &   1.16 &   3.1 &   0.5  &  Biggs (1992)        \\
J1933+2421 & B1931+24 &  0.814 &   8.11 &   1.6 &   80   &  Kramer et al. (2006)        \\
J1935+1616 & B1933+16 &  0.359 &   6.00 &   0.9 &   0.03 &  Biggs (1992)       \\
J1944+1755 & B1942+17 &  1.997 &   0.73 &  43.3 &   60    & Lorimer, Camilo \& Xilouris (2002)  \\
J1945$-$0040 & B1942$-$00 &  1.046 &   0.54 &  31.0 &   21   & Weisberg et al. (1986)        \\
J1946+1805 & B1944+17 &  0.441 &   0.02 & 290.0 &   55    &  Ritchings (1976)        \\
J2048$-$1616 & B2045$-$16 &  1.962 &  11.00 &   2.8 &   10  &  Ritchings (1976)        \\
J2113+4644 & B2111+46 &  1.015 &   0.71 &  22.5 &   12.5  &  Ritchings (1976)        \\
J2157+4017 & B2154+40 &  1.525 &   3.43 &   7.0 &   7.5  &  Ritchings (1976)        \\
J2305+3100 & B2303+30 &  1.576 &   2.89 &   8.6 &  10.0  &  Redman, Wright \& Rankin (2005)     \\
J2317+2149 & B2315+21 &  1.445 &   1.05 &  21.9 &  3.0  &  Weisberg et al. (1986)       \\
J2321+6024 & B2319+60 &  2.256 &   7.04 &   5.1 &   25    &  Ritchings (1976)        \\
J2330$-$2005 & B2327$-$20 &  1.644 &   4.63 &   5.6 &   12    &  Biggs (1992)        \\
\hline
\end{tabular} \\
Note: where Biggs (1992) states that nulls have been observed, but
only quotes an upper limit for the NF, the table value has been taken as half
the quoted upper limit.
\nocite{big92a} \nocite{dll+79} \nocite{fsk+04}
\nocite{htt70} \nocite{klo+06} \nocite{la83} \nocite{lcx02}
\nocite{lwf+04} \nocite{ran86} \nocite{rit76} \nocite{rwr05} \nocite{wab+86}
\end{minipage}
\end{table*}

\begin{figure}
\centerline{\psfig{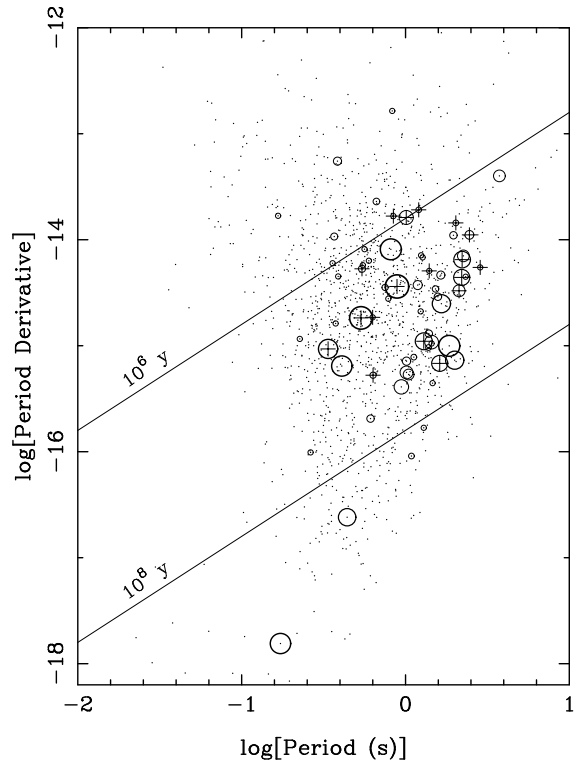}}
\caption{Plot of pulsar period versus period derivative for Galactic
  disk radio pulsars. Nulling pulsars are marked by a circle whose
  area is proportional to the null fraction with a lower
  bound of 5\%. Pulsars discussed in this paper are marked by a
  cross. Lines of constant characteristic age are shown.}\label{fg:ppdot}
\end{figure}

Fig.~\ref{fg:ppdot} shows the distribution of known nulling pulsars
on the $P$--$\dot P$ diagram.\footnote{Data from the ATNF Pulsar
  Catalogue (http://www.atnf.csiro.au/research/pulsar/psrcat;
  Manchester et al. 2005)}\nocite{mhth05} This diagram suggests that
the degree of nulling is related more to age than period. Although
some pulsars with large $\tau_c$ have small NF, all of the pulsars
with large NF have $\tau_c >$ 1 Myr and most have $\tau_c >$ 5 Myr. In
contrast, even excluding the recycled pulsar J1744$-$3922 with a period
of 0.172 s, PSR J1820$-$0509, with period of 0.337 s and age of 5.7
Myr has a NF of 67\%, whereas PSR B0525+21, with a period of 3.75 s
and an age of 1.5 Myr has a NF of 25\%. The longest-period pulsar in
our sample, PSR J1831$-$1223 (2.86 s), has a NF of just 4\%. 

There is no strong correlation of NF with profile morphology. As
Figs~\ref{fg:nulls} -- \ref{fg:1401} shows, nulls are observed in
pulsars with both narrow single-component pulse profiles and wider
multi-component profiles. As a fraction of the population, nulling
tends to be more common in multi-component profiles
\citep{ran86,big92a}; such pulsars tend to have larger $\tau_c$ so
this is just the same correlation as that shown in
Fig.~\ref{fg:ppdot}. Unfortunately, polarisation data exist for only a
few of these pulsars, so no clear relationships with impact parameter
or other polarisation characteristics can be drawn. There is no
obvious correlation of null cycle time with pulse period or
characteristic age.

Do RRATs represent an extreme form of pulse nulling or are they an
intrinsically different phenomenon? Based on a study of the pulse
modulation in PSR B0656+14, \citet{wsrw06} have suggested that RRATs
are highly modulated pulsars which occasionally emit very intense
pulses but which are too distant for the ``normal'' emission to be
detectable \citep[cf.][]{jr02}. Note that these are not ``giant''
pulses as observed in the Crab pulsar \citep[e.g.,][]{hkwe03}, other
young pulsars \citep{jr03} and some millisecond pulsars
\citep[e.g.,][]{kbm+06}, but simply very intense ``normal'' pulses. If
this is the correct explanation, then RRATs are not directly related
to nulling pulsars.

On the other hand, nulling and mode changing appear to be intimately
related. For example, in PSR J1701$-$3726 (Fig.~\ref{fg:mode}), the two
modes are always separated by a null interval. \citet{vkr+02} find
that PSR B0809+74 emits in a different mode after nulls and
\citet{rwr05} find that different modes in PSR B2303+30 have
different nulling properties. PSR J1703$-$4851 (Fig.~\ref{fg:mode})
appears to be similar with few or no nulls in the weaker but more
common mode. As mentioned in \S~1, weak emission with a
different pulse profile has been found in the ``null'' intervals of
PSR B0826$-$34 \citep{elg+05}. Very similar situations exist in PSR
J1326$-$6700 (Fig.~\ref{fg:1326}) where emission from the normal (mode
A) profile ceases and is replaced by emission from a new leading
component, and PSRs J1648$-$4458 and J1658$-$4306, where weak emission
is observed in what at first glance appear to be nulls. Simultaneous
multi-frequency observations \citep[e.g.,][]{bmsh82} show that both
nulling and mode changing are broad-band phenomena. Finally, there is
the fascinating observation by \citet{klo+06} of the change in slow-down
rate of PSR B1931+24 when the pulsar is in a null state.

All of these observations suggest that both nulling and mode changing
result from large-scale and persistent changes in the magnetospheric
current distribution.  Mode changes must be a manifestation of a
redistribution of current flow in the magnetosphere, resulting in
changes in the radio beam emission pattern and hence in the observed
pulse profile. Nulls may result from a cessation of (or at least a
large reduction in) the current as suggested by the observations of
PSR B1931+24 \citep{klo+06}, but may also result from a current
redistribution which leads to a beam pattern with little or no power
in our direction. This latter interpretation is favoured by the
increasing number of detections of weak emission, generally with a
different pulse profile, in apparent null intervals.  Both nulls
and mode changes are typically sudden transitions, occurring within
one pulse period, although exceptions apparently exist \citep{dchr86,lwf+04}.

The concept of ``tipping points'' is well known in the study of
non-linear complex systems, for example, in climate science. This is a
situation where a small perturbation can lead to a sudden change to a
different quasi-stable state. The transitions are fundamentally
related to positive feedback and can be reversible.\footnote{For an
  interesting discussion of the concept, see
  http://www.realclimate.org/index.php/archives/2006/07/runaway-tipping-points-of-no-return/.}
Evidently the pulsar pulse emission process in at least some pulsars
is subject to such instabilities. For example, if a small perturbation
in current flow results in a change in the accelerating potential or
magnetic field configuration which enhances the change, an instability
could develop. The observations imply changes in current flow which
are quite drastic, completely changing the observed pulse profile and
its polarisation and modulation characteristics, and in some cases
making the pulsar unobservable. What causes these instablities, why
they tend to be bistable and what determines their timescale are
currently unanswered questions. It is clear that they are most common
in old pulsars but there is a wide range in the observed properties of
pulsars which are subject to mode changing and nulling.

\section*{Acknowledgments}
The Parkes radio telescope is part of the Australia Telescope which is
funded by the Commonwealth Government for operation as a National
Facility managed by CSIRO. 


\begin{thebibliography}{}

\bibitem[\protect\citeauthoryear{Backer}{Backer}{1970a}]{bac70a}
Backer D.~C.,  1970a, Nature, 228, 1297

\bibitem[\protect\citeauthoryear{Backer}{Backer}{1970b}]{bac70}
Backer D.~C.,  1970b, Nature, 228, 42

\bibitem[\protect\citeauthoryear{Backer}{Backer}{1973}]{bac73}
Backer D.~C.,  1973, ApJ, 182, 245

\bibitem[\protect\citeauthoryear{Bartel, Morris, Sieber \& Hankins}{Bartel
  et~al.}{1982}]{bmsh82}
Bartel N.,  Morris D.,  Sieber W.,    Hankins T.~H.,  1982, ApJ, 258, 776

\bibitem[\protect\citeauthoryear{Biggs}{Biggs}{1992}]{big92a}
Biggs J.~D.,  1992, ApJ, 394, 574

\bibitem[\protect\citeauthoryear{Deich, Cordes, Hankins \& Rankin}{Deich
  et~al.}{1986}]{dchr86}
Deich W. T.~S.,  Cordes J.~M.,  Hankins T.~H.,    Rankin J.~M.,  1986, ApJ,
  300, 540

\bibitem[\protect\citeauthoryear{Durdin, Large, Little, Manchester, Lyne \&
  Taylor}{Durdin et~al.}{1979}]{dll+79}
Durdin J.~M.,  Large M.~I.,  Little A.~G.,  Manchester R.~N.,  Lyne A.~G.,
  Taylor J.~H.,  1979, MNRAS, 186, 39P

\bibitem[\protect\citeauthoryear{{Esamdin}, {Lyne}, {Graham-Smith}, {Kramer},
  {Manchester} \& {Wu}}{{Esamdin} et~al.}{2005}]{elg+05}
{Esamdin} A.,  {Lyne} A.~G.,  {Graham-Smith} F.,  {Kramer} M.,  {Manchester}
  R.~N.,    {Wu} X.,  2005, MNRAS, 356, 59

\bibitem[\protect\citeauthoryear{{Faulkner}, {Stairs}, {Kramer}, {Lyne},
  {Hobbs}, {Possenti}, {Lorimer}, {Manchester}, {McLaughlin}, {D'Amico},
  {Camilo} \& {Burgay}}{{Faulkner} et~al.}{2004}]{fsk+04}
{Faulkner} A.~J.,  {Stairs} I.~H.,  {Kramer} M.,  {Lyne} A.~G.,  {Hobbs} G.,
  {Possenti} A.,  {Lorimer} D.~R.,  {Manchester} R.~N.,  {McLaughlin} M.~A.,
  {D'Amico} N.,  {Camilo} F.,    {Burgay} M.,  2004, MNRAS, 355, 147

\bibitem[\protect\citeauthoryear{Gil, Jessner, Kijak, Kramer, Malofeev, Malov,
  Seiradakis, Sieber \& Wielebinski}{Gil et~al.}{1994}]{gjk+94}
Gil J.~A.,  Jessner A.,  Kijak J.,  Kramer M.,  Malofeev V.,  Malov I.,
  Seiradakis J.~H.,  Sieber W.,    Wielebinski R.,  1994, A\&A, 282, 45

\bibitem[\protect\citeauthoryear{{Hankins}, {Kern}, {Weatherall} \&
  {Eilek}}{{Hankins} et~al.}{2003}]{hkwe03}
{Hankins} T.~H.,  {Kern} J.~S.,  {Weatherall} J.~C.,    {Eilek} J.~A.,  2003,
  Nature, 422, 141

\bibitem[\protect\citeauthoryear{{Hobbs}, {Faulkner}, {Stairs}, {Camilo},
  {Manchester}, {Lyne}, {Kramer}, {D'Amico}, {Kaspi}, {Possenti}, {McLaughlin},
  {Lorimer}, {Burgay}, {Joshi} \& {Crawford}}{{Hobbs} et~al.}{2004}]{hfs+04}
{Hobbs} G.,  {Faulkner} A.,  {Stairs} I.~H.,  {Camilo} F.,  {Manchester} R.~N.,
   {Lyne} A.~G.,  {Kramer} M.,  {D'Amico} N.,  {Kaspi} V.~M.,  {Possenti} A.,
  {McLaughlin} M.~A.,  {Lorimer} D.~R.,  {Burgay} M.,  {Joshi} B.~C.,
  {Crawford} F.,  2004, MNRAS, 352, 1439

\bibitem[\protect\citeauthoryear{Huguenin, Taylor \& Troland}{Huguenin
  et~al.}{1970}]{htt70}
Huguenin G.~R.,  Taylor J.~H.,    Troland T.~H.,  1970, ApJ, 162, 727

\bibitem[\protect\citeauthoryear{{Janssen} \& {van Leeuwen}}{{Janssen} \& {van
  Leeuwen}}{2004}]{jl04}
{Janssen} G.~H.,  {van Leeuwen} J.,  2004, A\&A, 425, 255

\bibitem[\protect\citeauthoryear{Johnston, Lyne, Manchester, Kniffen, D'Amico,
  Lim \& Ashworth}{Johnston et~al.}{1992}]{jlm+92}
Johnston S.,  Lyne A.~G.,  Manchester R.~N.,  Kniffen D.~A.,  D'Amico N.,  Lim
  J.,    Ashworth M.,  1992, MNRAS, 255, 401

\bibitem[\protect\citeauthoryear{Johnston \& Romani}{Johnston \&
  Romani}{2002}]{jr02}
Johnston S.,  Romani R.,  2002, MNRAS, 332, 109

\bibitem[\protect\citeauthoryear{Johnston \& Romani}{Johnston \&
  Romani}{2003}]{jr03}
Johnston S.,  Romani R.,  2003, ApJ, 590, L95

\bibitem[\protect\citeauthoryear{Kaspi \& Stinebring}{Kaspi \&
  Stinebring}{1992}]{ks92}
Kaspi V.~M.,  Stinebring D.~R.,  1992, ApJ, 392, 530

\bibitem[\protect\citeauthoryear{{Knight}, {Bailes}, {Manchester}, {Ord} \&
  {Jacoby}}{{Knight} et~al.}{2006}]{kbm+06}
{Knight} H.~S.,  {Bailes} M.,  {Manchester} R.~N.,  {Ord} S.~M.,    {Jacoby}
  B.~A.,  2006, ApJ, 640, 941

\bibitem[\protect\citeauthoryear{{Kramer}, {Bell}, {Manchester}, {Lyne},
  {Camilo}, {Stairs}, {D'Amico}, {Kaspi}, {Hobbs}, {Morris}, {Crawford},
  {Possenti}, {Joshi}, {McLaughlin}, {Lorimer} \& {Faulkner}}{{Kramer}
  et~al.}{2003}]{kbm+03}
{Kramer} M.,  {Bell} J.~F.,  {Manchester} R.~N.,  {Lyne} A.~G.,  {Camilo} F.,
  {Stairs} I.~H.,  {D'Amico} N.,  {Kaspi} V.~M.,  {Hobbs} G.,  {Morris} D.~J.,
  {Crawford} F.,  {Possenti} A.,  {Joshi} B.~C.,  {McLaughlin} M.~A.,
  {Lorimer} D.~R.,    {Faulkner} A.~J.,  2003, MNRAS, 342, 1299

\bibitem[\protect\citeauthoryear{{Kramer}, {Lyne}, {O'Brien}, {Jordan} \&
  {Lorimer}}{{Kramer} et~al.}{2006}]{klo+06}
{Kramer} M.,  {Lyne} A.~G.,  {O'Brien} J.~T.,  {Jordan} C.~A.,    {Lorimer}
  D.~R.,  2006, Science, 312, 549

\bibitem[\protect\citeauthoryear{{Lewandowski}, {Wolszczan}, {Feiler},
  {Konacki} \& {So{\l}tysi{\' n}ski}}{{Lewandowski} et~al.}{2004}]{lwf+04}
{Lewandowski} W.,  {Wolszczan} A.,  {Feiler} G.,  {Konacki} M.,
  {So{\l}tysi{\' n}ski} T.,  2004, ApJ, 600, 905

\bibitem[\protect\citeauthoryear{{Lorimer}, {Camilo} \& {Xilouris}}{{Lorimer}
  et~al.}{2002}]{lcx02}
{Lorimer} D.~R.,  {Camilo} F.,    {Xilouris} K.~M.,  2002, ApJ, 123, 1750

\bibitem[\protect\citeauthoryear{Lorimer, Faulkner, Lyne, Manchester, Kramer,
  McLaughlin, Hobbs, Possenti, Stairs, Camilo, Burgay, D'Amico, Corongui \&
  Crawford}{Lorimer et~al.}{2006}]{lfl+06}
Lorimer D.~R.,  Faulkner A.,  Lyne A.~G.,  Manchester R.~N.,  Kramer M.,
  McLaughlin M.~A.,  Hobbs G.,  Possenti A.,  Stairs I.~H.,  Camilo F.,  Burgay
  M.,  D'Amico N.,  Corongui A.,    Crawford F.,  2006, MNRAS

\bibitem[\protect\citeauthoryear{Lyne \& Ashworth}{Lyne \&
  Ashworth}{1983}]{la83}
Lyne A.~G.,  Ashworth M.,  1983, MNRAS, 204, 519

\bibitem[\protect\citeauthoryear{McLaughlin, Stairs, Kaspi, Lorimer, Kramer,
  Lyne, Manchester, Camilo, Hobbs, Possenti, D'Amico \& Faulkner}{McLaughlin
  et~al.}{2003}]{msk+03}
McLaughlin M.~A.,  Stairs I.~H.,  Kaspi V.~M.,  Lorimer D.~R.,  Kramer M.,
  Lyne A.~G.,  Manchester R.~N.,  Camilo F.,  Hobbs G.,  Possenti A.,  D'Amico
  N.,    Faulkner A.~J.,  2003, ApJ, 591, L135

\bibitem[\protect\citeauthoryear{{Manchester}, {Hobbs}, {Teoh} \&
  {Hobbs}}{{Manchester} et~al.}{2005}]{mhth05}
{Manchester} R.~N.,  {Hobbs} G.~B.,  {Teoh} A.,    {Hobbs} M.,  2005, AJ, 129,
  1993

\bibitem[\protect\citeauthoryear{Manchester, Lyne, Camilo, Bell, Kaspi,
  D'Amico, McKay, Crawford, Stairs, Possenti, Morris \& Sheppard}{Manchester
  et~al.}{2001}]{mlc+01}
Manchester R.~N.,  Lyne A.~G.,  Camilo F.,  Bell J.~F.,  Kaspi V.~M.,  D'Amico
  N.,  McKay N. P.~F.,  Crawford F.,  Stairs I.~H.,  Possenti A.,  Morris
  D.~J.,    Sheppard D.~C.,  2001, MNRAS, 328, 17

\bibitem[\protect\citeauthoryear{{McLaughlin}, {Lyne}, {Lorimer}, {Kramer},
  {Faulkner}, {Manchester}, {Cordes}, {Camilo}, {Possenti}, {Stairs}, {Hobbs},
  {D'Amico}, {Burgay} \& {O'Brien}}{{McLaughlin} et~al.}{2006}]{mll+06}
{McLaughlin} M.~A.,  {Lyne} A.~G.,  {Lorimer} D.~R.,  {Kramer} M.,  {Faulkner}
  A.~J.,  {Manchester} R.~N.,  {Cordes} J.~M.,  {Camilo} F.,  {Possenti} A.,
  {Stairs} I.~H.,  {Hobbs} G.,  {D'Amico} N.,  {Burgay} M.,    {O'Brien} J.~T.,
   2006, Nature, 439, 817

\bibitem[\protect\citeauthoryear{{Morris}, {Hobbs}, {Lyne}, {Stairs}, {Camilo},
  {Manchester}, {Possenti}, {Bell}, {Kaspi}, {Amico}, {McKay}, {Crawford} \&
  {Kramer}}{{Morris} et~al.}{2002}]{mhl+02}
{Morris} D.~J.,  {Hobbs} G.,  {Lyne} A.~G.,  {Stairs} I.~H.,  {Camilo} F.,
  {Manchester} R.~N.,  {Possenti} A.,  {Bell} J.~F.,  {Kaspi} V.~M.,  {Amico}
  N.~D.,  {McKay} N.~P.~F.,  {Crawford} F.,    {Kramer} M.,  2002, MNRAS, 335,
  275

\bibitem[\protect\citeauthoryear{Rankin}{Rankin}{1986}]{ran86}
Rankin J.~M.,  1986, ApJ, 301, 901

\bibitem[\protect\citeauthoryear{{Redman}, {Wright} \& {Rankin}}{{Redman}
  et~al.}{2005}]{rwr05}
{Redman} S.~L.,  {Wright} G.~A.~E.,    {Rankin} J.~M.,  2005, MNRAS, 357, 859

\bibitem[\protect\citeauthoryear{Rickett}{Rickett}{1990}]{ric90}
Rickett B.~J.,  1990, Ann. Rev. Astr. Ap., 28, 561

\bibitem[\protect\citeauthoryear{Ritchings}{Ritchings}{1976}]{rit76}
Ritchings R.~T.,  1976, MNRAS, 176, 249

\bibitem[\protect\citeauthoryear{Taylor, Manchester \& Huguenin}{Taylor
  et~al.}{1975}]{tmh75}
Taylor J.~H.,  Manchester R.~N.,    Huguenin G.~R.,  1975, ApJ, 195, 513

\bibitem[\protect\citeauthoryear{{van Leeuwen}, {Kouwenhoven}, {Ramachandran},
  {Rankin} \& {Stappers}}{{van Leeuwen} et~al.}{2002}]{vkr+02}
{van Leeuwen} A.~G.~J.,  {Kouwenhoven} M.~L.~A.,  {Ramachandran} R.,  {Rankin}
  J.~M.,    {Stappers} B.~W.,  2002, A\&A, 387, 169

\bibitem[\protect\citeauthoryear{Weisberg, Armstrong, Backus, Cordes, Boriakoff
  \& Ferguson}{Weisberg et~al.}{1986}]{wab+86}
Weisberg J.~M.,  Armstrong B.~K.,  Backus P.~R.,  Cordes J.~M.,  Boriakoff V.,
    Ferguson D.~C.,  1986, AJ, 92, 621

\bibitem[\protect\citeauthoryear{{Weltevrede}, {Edwards} \&
  {Stappers}}{{Weltevrede} et~al.}{2006}]{wes06}
{Weltevrede} P.,  {Edwards} R.~T.,    {Stappers} B.~W.,  2006, A\&A, 445, 243

\bibitem[\protect\citeauthoryear{{Weltevrede}, {Stappers}, {Rankin} \&
  {Wright}}{{Weltevrede} et~al.}{2006}]{wsrw06}
{Weltevrede} P.,  {Stappers} B.~W.,  {Rankin} J.~M.,    {Wright} G.~A.~E.,
  2006, ApJ, 645, L149

\end{thebibliography}

\end{document}